\def \cm{~\rm{cm}}
\def \s{~\rm{s}}
\def \km{~\rm{km}}

\def \K{~\rm{K}}

\def \yr{~\rm{yr}}

\documentclass[vecphys]{svmult}

\usepackage{graphicx}        % standard LaTeX graphics tool
\usepackage[bottom]{footmisc}% places footnotes at page bottom

\begin{document}

\title*{Shaping Planetary Nebulae by Jets}
\author{Muhammad Akashi}
\institute{Department of Physics, Technion - Israel Institute of
Technology \texttt{akashi@physics.technion.ac.il}}

\maketitle

\begin{abstract}
We conduct 2D axisymmetrical hydrodynamical simulations to
investigate the interaction of a collimated fast wind (CFW; wide
jets) with a spherical AGB wind. The code includes radiative
cooling. We find that the shape of the planetary nebula (PN) is
sensitive to the exact mass loss history of the AGB wind, and the
opening angle of the CFW. Some typical PN morphologies are
obtained, but many other observed morphologies seem to require
more ingredients than what we assume in our present simulations,
e.g., equatorial AGB wind, and ionization and fast wind during the
PN phase. The hot bipolar bubble formed by the jets is an X-ray
source. 
\end{abstract}

\section{Numerical Simulations}
\label{sec:numeric}
%\label{sec:1}
% Always give a unique label
% and use \ref{<label>} for cross-references
% and \cite{<label>} for bibliographic references
% use \sectionmark{}
% to alter or adjust the section heading in the running head
%Your text goes here. Use the \LaTeX\ automatism for your citations
%\cite{monograph}.

Our simulations were performed using Virginia Hydrodynamics-I
(VH-1), a high resolution multidimensional astrophysical
hydrodynamics code developed by John Blondin and co-workers
(Blondin et al. 1990; Stevens et al., 1992; Blondin 1994). We have
added radiative cooling to the code at all temperatures $T >
T_{\rm min}$, where the gas temperature is bound from below at
$T_{\rm min}=300-1000 \K$. Radiative cooling is carefully treated
near contact discontinuities, where large temperature gradients
can lead to unphysical results.

We simulate axisymmetrical morphologies. This allows us to use
axisymmetrical grid, and to simulate one quarter of the meridional
plane. There are 208 grid points in the azimuthal ($\theta$)
direction of this one quarter and 208 grid points in the radial
direction. The radial size of the grid points increases with
radius. In these simulations the grid extends from $10^{15} \cm$
to $4 \times 10^{17} \cm$.

Before the jet is launched ($t=0$), the grid is filled with slow
wind having a speed $v_1=10 \km \s^{-1}$  and mass loss rate $\dot
M_1 \sim 10^{-5} M_\odot \yr^{-1}$, with small variations between
the runs. We launch a collimated fast wind from the first 20 zones
attached to the inner boundary of the grid. The jet is uniformly
ejected within an angle (half opening angle) $\alpha$ ($ 0 \le
\theta \le \alpha$). For numerical reasons a weak slow wind is
injected in the sector $ \alpha < \theta \le 90^\circ$ ($\theta =
0$ along the symmetry axis - vertical in the figures here).

%================================================================================
\begin{figure}
\centering
\resizebox{0.59\textwidth}{!}{\includegraphics{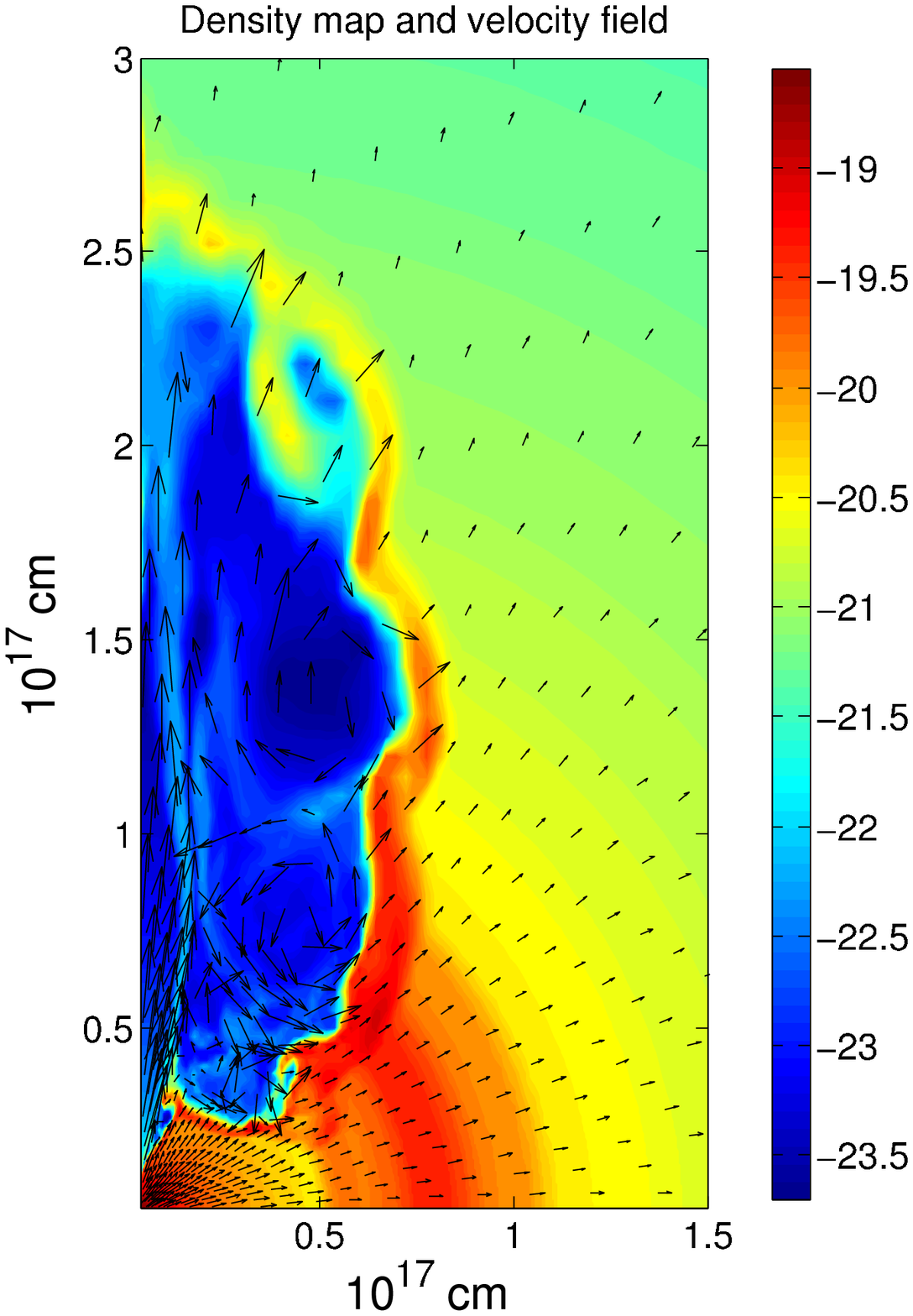}}
\centering
\resizebox{0.89\textwidth}{!}{\includegraphics{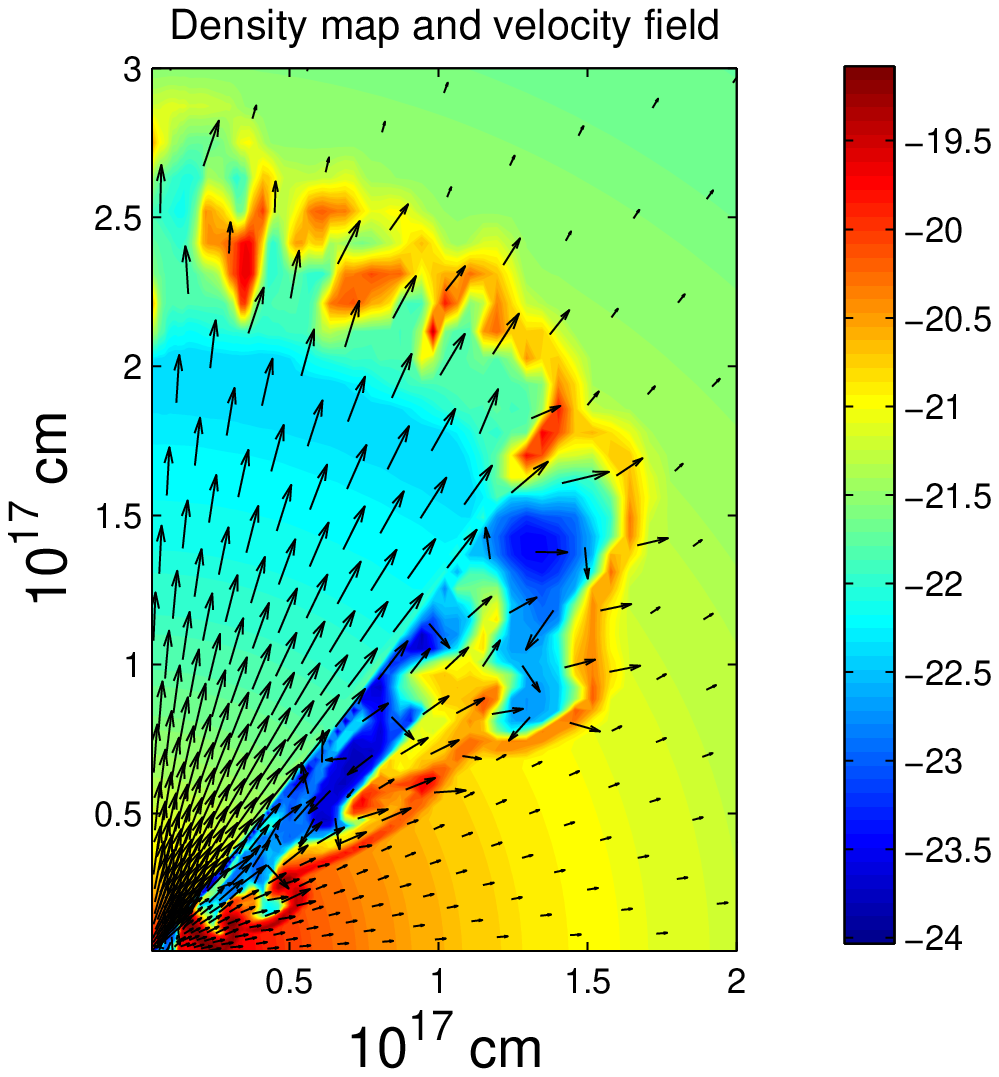}}

\caption{Top: Density plot at  $t = 2000 \yr$ (after jet launching
starts) for a case where the slow wind mass loss rate has
increased by a factor of five to $\dot M_1 = 5 \times 10^{-5}
M_\odot \yr^{-1}$ $950 \yr$  before jet starts. The vertical axis
is the symmetry axis and the horizontal axis is in the equatorial
plane. Slow spherical wind: at $t=0$ $v_1=10 \km \s^{-1}$, $\dot
M_1 = 5 \times 10^{-5} M_\odot \yr^{-1}$ for $r < 3 \times 10^{16}
\cm$, and $\dot M_1 = 10^{-5} M_\odot \yr^{-1}$ for $r > 3 \times
10^{16} \cm$. The jet has a half opening angle  $\alpha=20^\circ$,
$v_2=600 \km \s^{-1}$, and $\dot M_2 = 2 \times 10^{-8} M_\odot
\yr^{-1}$. Arrows indicate flow direction: $v > 200 \km \s^{-1}$
(long arrow), $ 20 < v \le 200 \km \s^{-1}$ (medium arrow), and $
v \le 20 \km \s^{-1}$ (short arrow). This general case can
account for PNs such as NGC 6886 or NGC 3918. Bottom: Density plot
at $t = 530 \yr$ for another case. Slow spherical wind: $v_1=10
\km \s^{-1}$, $\dot M_1 = 5 \times 10^{-6} M_\odot \yr^{-1}$. The
jet has a half opening angle $\alpha=40^\circ$, $v_2=600 \km
\s^{-1}$, and $\dot M_2 = 2.5 \times 10^{-6} M_\odot \yr^{-1}$.
This general case can account for some morphological features in
He 2-104, such as the instabilities in the outer lobes.}
\label{neb1}
\end{figure}

%================================================================================
%\begin{figure}
%\centering
%\resizebox{0.59\textwidth}{!}{\includegraphics{Pakashif2.eps}}

%\picplace{5cm}{2cm} % Give the correct figure height and width in cm
%
%\caption{Density plot at  $t = 1300 \yr$ for a case with a slow wind
%of $v_1=10 \km \s^{-1}$ , and mass loss rate of $\dot M_1 = 3 \times 10^{-5} M_\odot \yr^{-1}$
%for $r < 3 \times 10^{16} \cm$, and $\dot M_1 = 5 \times 10^{-6} M_\odot \yr^{-1}$
%for $r > 3 \times 10^{16} \cm$.
%The jet has a half opening angle  $\alpha=30^\circ$, $v_2=600 \km \s^{-1}$ ,
%and  $\dot M_2 = 10^{-7} M_\odot \yr^{-1}$  .
%The initial temperature was 1000 \K, and the temperature is limited from below at 1000 \K.
%The arrows are just as in the previous figure.}
%\end{figure}

%================================================================================
\begin{figure}
\centering
\resizebox{0.59\textwidth}{!}{\includegraphics{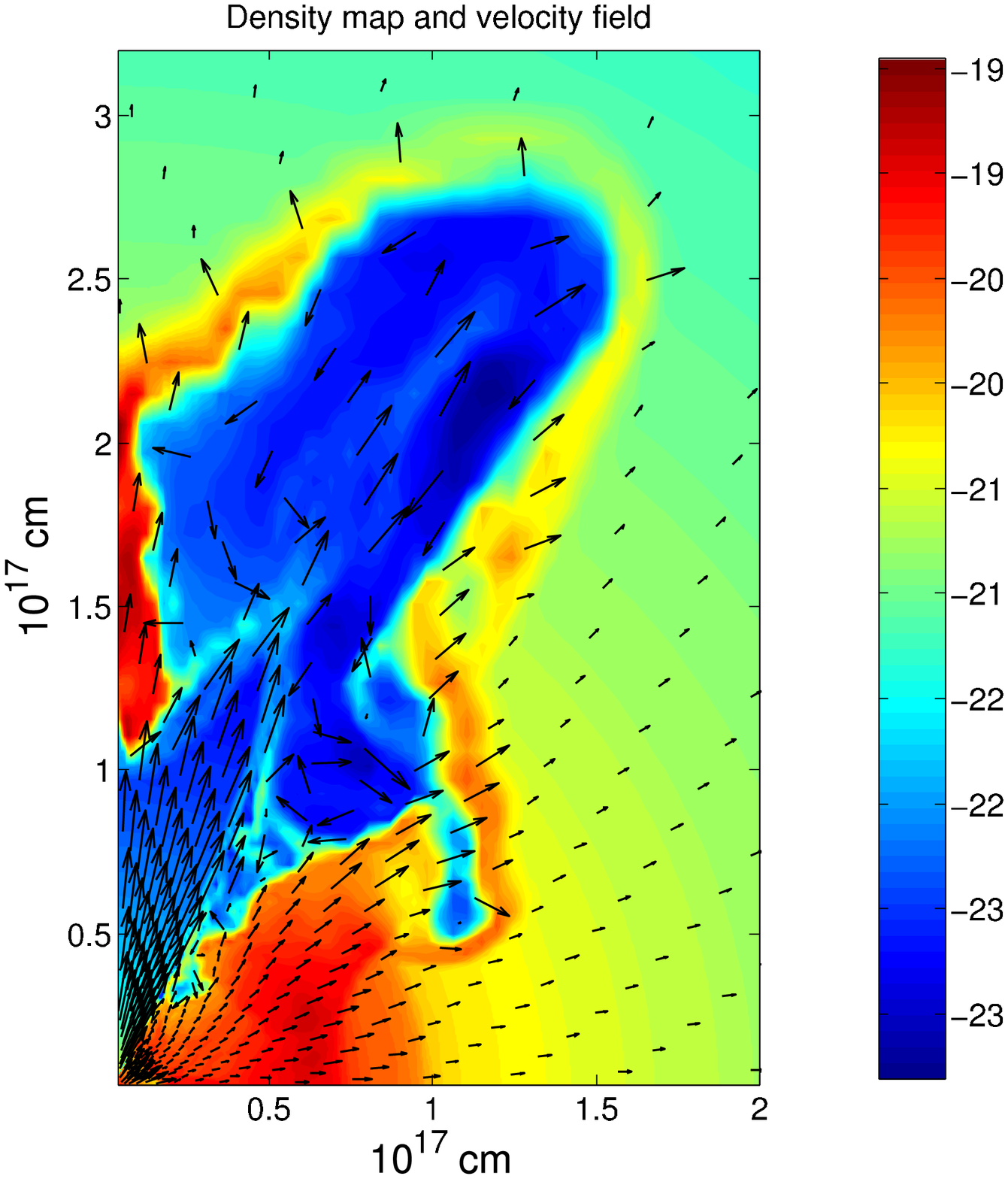}}
\centering
\resizebox{0.99\textwidth}{!}{\includegraphics{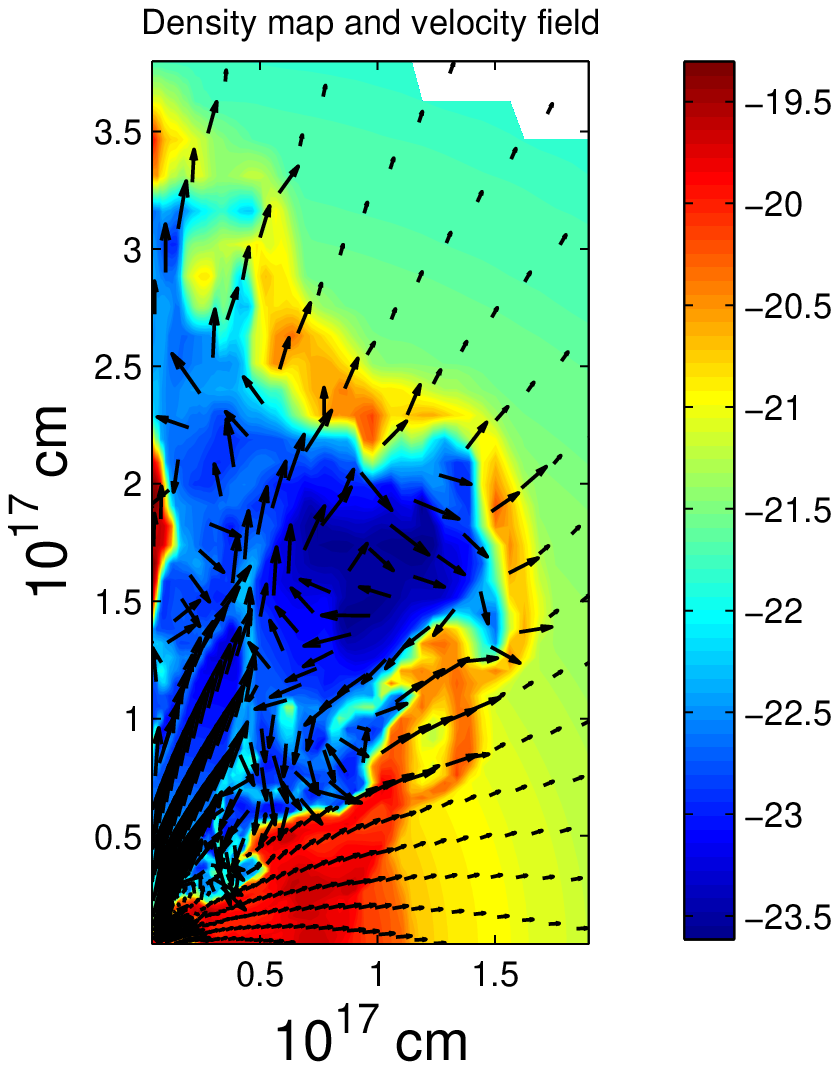}}
%\picplace{5cm}{2cm} % Give the correct figure height and width in cm
%
\caption{Top: Density plot at $t = 1300 \yr$ for a case with a
slow wind of $v_1=10 \km \s^{-1}$, and mass loss rate of $\dot M_1
= 3 \times 10^{-5} M_\odot \yr^{-1}$ for $r < 3 \times 10^{16}
\cm$, and $\dot M_1 = 5 \times 10^{-6} M_\odot \yr^{-1}$ for $r >
3 \times 10^{16} \cm$ at $t=0$. The jet has a half opening angle
$\alpha=30^\circ$, $v_2=600 \km \s^{-1}$, and  $\dot M_2 = 10^{-7}
M_\odot \yr^{-1}$. The initial temperature was 1000 \K, and the
temperature is limited from below at 1000 \K. Bottom: Density plot
at $t = 1560 \yr$  for a case with a slow wind of $v_1=10 \km
\s^{-1}$, and mass loss rate of $\dot M_1 = 3 \times 10^{-5}
M_\odot \yr^{-1}$ for $r < 3 \times 10^{16} \cm$, and $\dot M_1 =
5 \times 10^{-6} M_\odot \yr^{-1}$ for $r > 3 \times 10^{16} \cm$
at $t=0$. The jet has a half opening angle $\alpha=30^\circ$,
$v_2=600 \km \s^{-1}$, and  $\dot M_2 = 10^{-7} M_\odot \yr^{-1}$.
The initial temperature was 10000 \K, and the temperature is
limited from below at 1000 \K. The general shape is similar to MZ
3. The arrows are just as in figure \ref{neb1}. Note that top and
bottom are the same beside the initial temperature.}
%\label{fig:3}       % Give a unique label
\label{neb2}
\end{figure}

%================================================================================

%\begin{figure}
%\centering
%\resizebox{0.59\textwidth}{!}{\includegraphics{Pakashif4.eps}}
%\caption{Density plot at  $t = 1560 \yr$  for a case with a slow wind
%of $v_1=10 \km \s^{-1}$, and mass loss rate of $\dot M_1 = 3 \times 10^{-5} M_\odot \yr^{-1}$
%for $r < 3 \times 10^{16} \cm$, and $\dot M_1 = 5 \times 10^{-6} M_\odot \yr^{-1}$
%  for $r > 3 \times 10^{16} \cm$.
%The jet has a half opening angle  $\alpha=30^\circ$, $v_2=600 \km \s^{-1}$ ,
%and  $\dot M_2 = 10^{-7} M_\odot \yr^{-1}$  .
%The initial temperature was 10000 \K, and the temperature is limited from below at 1000 \K.
%The general shape is similar to MZ 3. The arrows are just as in the previous figure.}
%\label{fig:3}       % Give a unique label
%\end{figure}
%================================================================================
\section{Results and Summary}

We try  to explain PN shaping and formation processes by
simulating jets interacting with the AGB wind. The parameter space
for these types of flows is huge. We run more than 200
simulations, in which we tried many values of the CFW (jets) mass
loss rate, velocity, and opening angle.

Another parameter that was used is the mass loss rate history of
the AGB wind. For example, in a model presented here (figure
\ref{neb2}), we assume that some period $\Delta t$  before the
beginning of the jet-launching phase the slow wind mass loss rate
increased by some factor $k$; e.g., in the run presented in figure
\ref{neb2} $\Delta t=950 \yr$ and $k=6$. We found that the mass
loss history of the AGB wind plays a significant role in
the shaping process. In some runs the AGB wind initial temperature
was set to $10,000 \K$ instead of $1,000 \K$, with a noticeable
influences on the final shape of the PN (compare the two panels in
figure \ref{neb2}). Taking the AGB wind to be at $\sim 10,000 \K$
is relevant to system where the accreting companion is a WD. If
the accretion rate is high enough the WD sustains a constant
nuclear burning, and it is very hot and luminous, as super-soft
X-ray sources (e.g., Starrfield et al. 2004). Such an accreting WD
might maintain the entire AGB wind ionized.

We were able to reproduce some physical properties of PNs, but not
all desirable properties. It is very difficult, and might be
impossible, to get the entire range of properties with our limited
numerical code. We must add more ingredients to the code. Examples
are pulsed jets, precessing jets, dense equatorial outflow and
ionization fronts at later times.

We presented results of four numerical runs that migh match real
PNs. Our main results are as follows:

\begin{enumerate}
\item We show that different PNs morphologies could be strongly
dependent on the AGB wind mass loss rate history. \item We have an
evident that the final morphology is very sensitive also to the
assumed initial temperature. For $T_{i}=1000 \K$ and $T_{i}=10000
\K$ we get two different evolutions of the planetary nebula (figure \ref{neb2}).
\item The dense finger along the symmetry axis (in
figure \ref{neb2}, top) result from instabilities. In reality, it
would be more extended and not so narrow. It is forced to the
symmetry axis by the numerical code. \item In one case we obtain
low density finger (in figure \ref{neb2}) protruding to the upper
right form a torus. However, it is also a result of an
instability, and we expect that in real systems there will be
several such fingers. In particular, if there is a departure from
axisymmetry due to the orbital motion, we would expect the fingers
to be similar but not identical in the two sides of the equatorial
plane. \item In all cases that we run we saw that there is a
strong dependence on the half opening angle of the jet. This will
be the focus of a future paper.
\end{enumerate}

\section{Acknowledgements}
I would like to acknowledge Noam Soker for his immense help with
understanding the numerical simulations and for his help for
getting the important results. I also thank John Blondin for his
help with the numerical code. This research was supported by the
Asher Space Research Institute.

%
%
% Use the following syntax and markup for your references
%

%%%%%%%%%%%%%%%%%%%%%%%%%%%%%%%%%%%%%%%%%%%%%%%%%%%%%%%%%%%%%%%%%%%%%%  }

%%%%%%%%%%%%%%%%%%%%%%%%%%%%%%%%%%%%%%%%%%%%%%%%%%%%%%%%%%%%%%%%%%%%%%

\end{document}